\documentclass[useAMS,usenatbib]{mn2e}
\usepackage{aas_macros}
\usepackage{amssymb}
\usepackage{amsmath}
\usepackage{graphicx}
\usepackage{dcolumn}
\usepackage{bm}
\usepackage{multicol}
\usepackage{subfig}
\usepackage{caption}
\usepackage[usenames]{color}
\usepackage{float}
\usepackage{psfrag}
\usepackage{fixltx2e}
\usepackage[mathscr]{euscript}

\newcommand{\alert}{\textcolor{black}}
\newcommand{\alerta}{\textcolor{black}}

\title[Bias in the X-ray vs SZ pressure profiles]{Bias from gas inhomogeneities in the pressure profiles as measured from X-ray and SZ observations}

\author[Khedekar et al.]{
\parbox[t]{16cm}{
S. Khedekar$^{1}$\thanks{satej@mpa-garching.mpg.de}, E. Churazov$^{1,2}$, A. Kravtsov$^{3,4}$, I. Zhuravleva$^{1,5}$, E.T. Lau$^{6,7}$, D. Nagai$^{6,7}$, R. Sunyaev$^{1,2}$
}\\
\\
$^{1}$MPI f\"ur Astrophysik, Karl-Schwarzschild str. 1, Garching, 85741, Germany\\
$^{2}$Space Research Institute, Profsoyuznaya str. 84/32, Moscow, 117997, Russia\\
$^{3}$Department of Astronomy and Astrophysics, University of Chicago, 5640 South Ellis Avenue, Chicago, IL 60637, USA\\
$^{4}$Kavli Institute for Cosmological Physics and Enrico Fermi Institute, University of Chicago, Chicago, IL 60637, USA\\
$^{5}$Kavli Institute for Particle Astrophysics and Cosmology, Stanford University, 452 Lomita Mall, Stanford, CA 94305-4085, USA\\
$^{6}$Department of Physics, Yale University, New Haven, CT 06520, USA\\
$^{7}$Yale Center for Astronomy and Astrophysics, Yale University, New Haven, CT 06520, USA\\
}

\begin{document}

\date{Draft: \today}

\pagerange{\pageref{firstpage}--\pageref{lastpage}} \pubyear{2012}
 
\maketitle

\label{firstpage}

\begin{abstract}
X-ray observations of galaxy clusters provide emission measure weighted spectra, arising from a range of density
 and temperature fluctuations in the intra-cluster medium (ICM). This is fitted to a single temperature plasma
 emission model to provide an estimate of the gas density and temperature, which are sensitive to the gas inhomogeneities.
 Therefore, X-ray observations yield a potentially biased estimate of the thermal gas pressure, $P_{X}$. At the same time
 Sunyaev-Zeldovich (SZ) observations directly measure the integrated gas pressure, $P_{SZ}$. If the X-ray pressure
 profiles are strongly biased with respect to the SZ, then one has the possibility to probe the gas inhomogeneities (their
 amplitude and physical nature), even at scales unresolved by the current generation of telescopes. At the same time,
 a weak bias has implications for the \alert{interchangeable} use of mass proxies like $Y_{SZ}$ and $Y_X$ as cosmological probes. In this
 paper we investigate the dependence of the bias, defined as $\displaystyle b_P(r)\equiv {P_X(r)}/{P_{SZ}(r)}-1$,
 on the characteristics of fluctuations in the ICM taking into account the correlation between temperature and density
 fluctuations. We made a simple prediction of the {\it irreducible} bias in idealised X-ray vs SZ observations
 using multi-temperature plasma emission model. We also provide a simple fitting form to estimate the bias given the
 distribution of fluctuations. In real observations there can be additional complications arising from
instrumental background, insufficient photon statistics, asphericity, method of deprojection, etc.
 Analysing a sample of 16 clusters extracted from hydrodynamical simulations, we find that the median value of
 bias is within $\pm3\%$ within $R_{500}$, it decreases to $-5\%$ at $R_{500}< r < 1.5~ R_{500}$
 and then rises back to $ \sim 0\%$ at $r \gtrsim 2~R_{500}$. The scatter of $b_P(r)$ between individual relaxed clusters
 is small -- at the level of $<0.03$ within $R_{500}$, but turns significantly larger (0.25) and highly skewed
 (${\overline b_P}(r) \gg 0$) at $r\gtrsim 1.5 ~R_{500}$. \alert{For any relaxed cluster we find $|b_P(r)| < 15\%$
 within $R_{500}$, across different implementations of input physics in the simulations. 
 Unrelaxed clusters \alert{exhibit} a larger scatter \alert{in $b_P(r)$} (both from radius to radius and from cluster to cluster).}
\end{abstract}


\section{Introduction}
Galaxy clusters offer an interesting possibility of probing physics relevant on both cosmological as well as galactic scales \citep[see recent
 review by][]{2012ARA&A..50..353K}. On the one hand, they are useful
 for constraining cosmological parameters \citep{2001ApJ...553..545H}, models of dark energy \citep{MM04, 2009ApJ...692.1060V, 2012AstL...38..347B}, and
 possible modifications to the theory of gravity \citep{2009MNRAS.400..699R}. At the same time, they offer insights into feedback processes
 involving the interplay between the intra-cluster medium (ICM) and constituent galaxies, and growth of supermassive black holes occurring at their centres
\alert{\citep[e.g.,][]{Churazov, Croton, Rafferty, Somerville}}.

Galaxy clusters emit strongly in X-rays from the shock heating and compression of the infalling matter \citep{1972A&A....20..189S}.
  In the last two decades, a large number clusters have been imaged at high resolution through X-ray instruments such as ROSAT,
  XMM-Newton and Chandra. The X-ray satellite, Suzaku has also been useful at imaging the outskirts of clusters \citep[e.g.,][]{Simionescu},
  due to its lower instrumental background. On the cosmic microwave background (CMB) sky, clusters are visible due to the inverse Compton
  scattering of the CMB photons by the hot ICM, known commonly as the SZ effect \citep{1972CoASP...4..173S}. In recent years ongoing SZ surveys
  like SPT \citep{2012arXiv1203.5775R}, ACT \citep{2011ApJ...737...61M}, Planck \citep{2011A&A...536A...8P} and BOLOCAM \citep{2012arXiv1211.1632S}
  have been imaging clusters on arc-minute scales using multi-frequency data.
 These surveys should detect $\gtrsim 1000$ clusters up to high redshifts making them valuable as cosmological probes. However, in order to place
 competitive constraints from these observations, the cluster masses would have to be determined with an accuracy of $\lesssim 5\%$ \citep{Allen}.

Joint observations of clusters in SZ, X-ray and optical bands provide multiple mass proxies such as integrated SZ flux, $Y_{SZ}$ \citep{Motl, NagaiGF};
 $Y_X = M_{\rm gas}T_X$ \citep{2006ApJ...650..128K} from X-ray observations; and measurements of velocity dispersions in the optical \citep{2008ApJ...672..122E}.
Forthcoming missions like ASTRO-H would probe the gas motions via X-ray spectroscopy \citep{2003AstL...29..791I, 2012MNRAS.422.2712Z}
 and help provide unbiased hydrostatic mass estimates \citep[e.g.,][]{2009ApJ...705.1129L}.
 Such multi-wavelength observations would be useful to check for the consistency of the derived masses and help in identifying the sources of scatter in the
 mass-observable scaling relations \citep{2012ApJ...758...74B, 2012arXiv1204.1577N}. Accurate knowledge of this scatter is important for precise determination
 of the cosmological parameters using cluster surveys \citep{2005PhRvD..72d3006L, Gladders}.

In this context, some of the recent efforts have focused on following up many of the SZ detected clusters in X-rays in order to probe the relationship between
 $Y_X$ and $Y_{SZ}$ \citep{2011ApJ...738...48A, 2011arXiv1112.5595P}, both of which measure the total thermal energy in the cluster. These works have reported
 the ratio $Y_{SZ}/Y_X$ at values $0.82\pm 0.07$ and $0.95\pm 0.03$ respectively, for measurements within a cluster radius of $R_{500}$\footnote{\alert{$R_{500}$
 is the radius, $r$, of the cluster within which the enclosed mass, $M(<r)$, equals, $M_{500} \equiv 500 \rho_{\rm crit}(z) \frac{4}{3} \pi r^3$;
 where $\rho_{\rm crit}(z)$ is the critical density of the universe at the observed redshift, $z$.}}. More recently,
 \citet{2012arXiv1202.2150R} independently reported this ratio to be $0.82\pm0.024$ using cluster data from Planck and Chandra observations. More data on joint
 observations of clusters would reduce the statistical and systematic uncertainties and determine the bias between $Y_X$ and $Y_{SZ}$ with better accuracy.

The deprojection analysis of X-ray surface brightness, first implemented for clusters by \citet{1981ApJ...248...47F}, is being routinely performed
 \citep[in various versions, for e.g.,][] {1983ApJ...272..439K, 1999AJ....117.2398M, 2006ApJ...640..691V} to yield the
 three-dimensional radial profiles of both temperature and density. Henceforth we shall use the term `deprojection' to imply the particular method
 as implemented in \citet{2003ApJ...590..225C} \citep[see also][]{2008MNRAS.390.1207R}. Briefly, assuming spherical symmetry, a set of measured
 values of the surface brightness in concentric rings is converted into a set of emissivities in spherical shells, using the inverse of a simple
 geometrical projection matrix. The procedure is repeated for many energies, yielding for each shell an energy dependent volume emissivity - i.e. the
 spectrum, which is then fitted in XSPEC with the standard model(s).
 
Thanks to rapid technological progress in SZ observations it has been possible to apply similar methods to well-resolved SZ images obtained with the Planck
 \citep{2012arXiv1208.3611P}, SPT \citep{SPT}, ACT \citep{ACT}, CARMA \citep{CARMA}, MUSTANG \citep{MUSTANG} and BOLOCAM \citep{2012arXiv1211.1632S}.
 From these measurements one may then obtain the volume integrated thermal pressure $\int P dV = \int n_e T dV$, in each concentric shell both in X-rays
 and SZ, following the deprojection method as outlined in the above paragraph.

\begin{figure*}
  \vspace*{-4.2cm}
  \hspace*{1.5cm}
  \centering
  \includegraphics[width=0.67\textwidth, angle=-90]{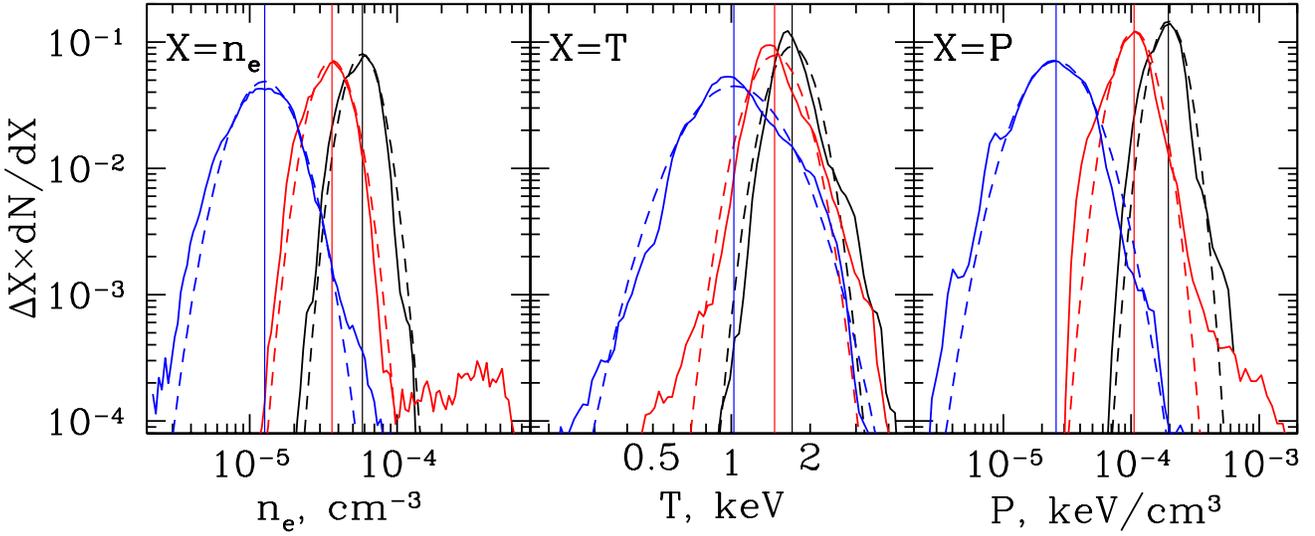}
  \caption{Distribution of density, temperature and pressure in simulated clusters \citep[see][for details]{Zhuravleva}. The solid lines show the actual
  distribution, while the dashed lines indicate the lognormal distribution having the same full width at half maxima. The
  distributions shown in blue/red/black colours are measured for the (highly relaxed) cluster CL7 in the CSF simulation
  (see section \ref{sec:simulations})
  in concentric shells at the radii, $r/R_{500}$: 0.9--1.0/1.1--1.2/1.6--1.8. Note the presence of an extended high density tail in the
  distribution of density fluctuations.}
  \label{fig:distrP}
\end{figure*}

 X-ray observations reveal the presence of substructures, even in relaxed clusters, with the fluctuations in density and temperature having amplitudes
 of $\sim$ 10\% \citep{2009ApJ...699.1178Z} at $r \lesssim 0.5~ R_{500}$ \citep[see also][]{2012MNRAS.421.1123C}.
 Zhuravleva et al. (2012b), henceforth Z12, investigated properties of the ICM fluctuations in simulated clusters and found their distribution
 to be near-lognormal along with a high density tail \alerta{contributing $\sim$ 1\% by volume \citep[see also,][]{Roncarelli, Vazza}}.
The width of the density fluctuations is small (0.1 dex) at the centre of relaxed clusters but grows by a factor of 6 at $2~R_{500}$; with unrelaxed clusters
 found to have a broader distribution than relaxed clusters. Z12 propose a simple method for identifying the high density clumps using a
 sample of simulated clusters and show that the median values of the ICM pressure remain robust to the extraction of such clumps.

If we assume that the X-ray emissivity is independent of temperature\footnote{The X-ray emissivity in the
energy band $E1$--$E2$ is $\epsilon_X \propto n_e^2 \Lambda(E1,E2,T)$. For $T \gtrsim 3$ keV the soft band (0.5 -- 2 keV)
emissivity is almost independent of the temperature.}, and the width of the lognormal distribution of gas density fluctuations,
$\sigma_{ne}$ is small ($\sigma_{ne} \ll 1$), the bias, \alert{$b_P$,} between the pressures measured from X-ray \alert{($P_X$)} and SZ \alert{($P_{SZ}$)} observations,
\alert{as a function of $\sigma_{ne}$, in the following three different hypotheses (isothermal, adiabatic and isobaric gas) may be written as,}

{\setlength{\arraycolsep}{2pt}
\begin{eqnarray}
  \label{eqn:estimate}
    b_P &\equiv& \frac{P_X}{P_{SZ}} -1 =  \frac{\sqrt{{\langle n_e^2 \rangle}}\langle T \rangle}{\langle n_e T \rangle} -1  \\
        &=& \exp\Big[\frac{\sigma_{ne}^2}{2}\Big]-1 \approx \frac{\sigma_{ne}^2}{2}~~~~~~~~~~~~~~~ \dots~ {\rm \it isothermal } \nonumber \\
        &=& \exp\Big[\Big(\frac{3}{2}-\gamma\Big)\sigma_{ne}^2\Big]-1 \approx -\frac{\sigma_{ne}^2}{6} ~~\dots~{\rm \it adiabatic} \nonumber \\
        &=& \exp\Big[\frac{3\sigma_{ne}^2}{2}\Big]-1 \approx \frac{3}{2}\sigma_{ne}^2~~~~~~~~~~~~ \dots~ {\rm \it isobaric } \nonumber
\end{eqnarray}}for the adiabatic index, $\gamma=5/3$. \alert{It is interesting to note here that the bias takes positive values for isobaric fluctuations
 but is negative for adiabatic fluctuations.}

Thus, for such fluctuations one expects the bias to be simply related to the gas clumping factor,
\mbox{$C \equiv {\langle n_e^2 \rangle}/{\langle n_e \rangle}^2 = \exp(\sigma_{ne}^2) \approx 1 + \sigma_{ne}^2$}.
 Note also that the magnitude of this bias is small as long as the fluctuations are not large.
 \citet{1999ApJ...520L..21M} showed using simulations that the average bias in gas mass within $R_{500}$ due to clumpiness in the ICM is about +16\%,
 while \citet{Nagai2007b} found the bias to be lower at $\lesssim 6$\%. \citet{2011ApJ...731L..10N} showed that clumping may be even
 larger ($C \sim 2-6$) at the cluster outskirts ($r \gtrsim 1-1.5~ R_{500}$), \citep[see also][]{2012arXiv1211.1695V}.

X-ray and SZ observations provide complimentary information for studying clumpiness and fluctuations in the ICM. However,
 in reality the actual bias between $P_X$ and $P_{SZ}$ may well be different from the simple estimate presented in equation \ref{eqn:estimate}
 due to the effects of X-ray instrumental response and, more importantly, the bias resulting from fitting a single temperature
 model to a spectrum produced by gas with a range of temperatures \citep[e.g.][]{2004MNRAS.354...10M, 2007ApJ...659..257K}.
 For example, in \alerta{a uniform} gas having only temperature fluctuations, the measured temperature is weighted down by
 $\Lambda(E1,E2,T)$ due to the relative importance of line spectra at lower temperatures, producing a negatively biased $P_X$ with
 respect to $P_{SZ}$. In addition, there might also be a bias arising from the application of a simple $\beta$ model \citep{2006MNRAS.369.2013R}
 or from the assumption of spherical symmetry for clusters having more complicated morphology \citep{2007MNRAS.382..397A}, however we shall
 not discuss these issues here.

 It is therefore necessary to make a realistic estimate of the {\it irreducible} bias between X-ray and SZ measurements of pressure profiles.
 By irreducible, we mean the bias that would persist even for the most favourable or idealised observations. \alert{In this work, we estimate
 this bias in two ways: (i) From an idealised hot gas described by a lognormal distribution of density and temperature values; (ii) and then
 from \alerta{hydrodynamical simulations} (with different implementations of input physics) of clusters in a broad mass range. Mock X-ray spectra
 generated from the given distributions of gas density and temperature are used to estimate this bias.}
Our computation of this bias takes into account both the density and temperature fluctuations along with their correlation, as well as the effects
 of the X-ray instrumental response function (for the Chandra telescope) involved in the spectral fitting.
We wish to emphasise that in real observations, the bias is likely to depend on the method of deprojection, X-ray instrumental background, ability
 to remove dense clumps from images (which may be limited by photon statistics), etc. The actual bias inherent in a given X-ray/SZ analysis
 method should be estimated via detailed tests using the analysis method outlined in section \ref{sec:method} on the mock X-ray observations.

 \alert{The remainder of the paper is organised as follows: In section \ref{sec:method}, we describe the method used to compute the pressure bias from
 X-ray and SZ observations and apply it to an idealised gas described by a bivariate lognormal distribution of density and temperature fluctuations.
 In section \ref{sec:simulations} we briefly describe the simulations that were used to estimate
 the bias. \alerta{In section \ref{sec:results} we discuss the properties of density and temperature fluctuations in
 the ICM of simulated clusters and define} two procedures used to remove the contribution of high density clumpy regions.
 Next we present the radial profiles of bias in X-ray/SZ pressure measurements along with the biases in the observed densities and temperatures from X-ray
 observations. In section \ref{sec:compare}, we compare our results with current and future observations and also with some recent results presented
 in other theoretical works. Lastly we highlight our main results and conclude.}

\section{\alert{The Bias from an idealised hot gas with a lognormal distribution}}
\label{sec:method}
 Hydrodynamical simulations of clusters show
 that the density, $n_e$, and temperature, $T$, of the ICM follow a nearly lognormal distribution \citep{2007ApJ...659..257K, Zhuravleva}
 with an extended tail of high density fluctuations
(HDF), see Fig. \ref{fig:distrP}.  We model an idealised hot gas ($T \sim {\rm \it few} \ {\rm keV}$) as a weighted sum of 30 $\times$ 30 components (in densities
 and temperature) with the weights (here, the emission measures) correctly normalised to follow a bivariate lognormal distribution in $n_e$
 and $T$, defined by the following probability density function,
\begin{eqnarray}
  \label{eqn:lognormal}
  \displaystyle
  f(n_e, T) = \frac{1}{2\pi n_e T \lvert {\boldsymbol {\mathcal C}} \rvert} \exp{\left(-\frac{1}{2} {{\boldsymbol X} \cdot {\boldsymbol {\mathcal C}}^{-1} \cdot {\boldsymbol X}^T} \right)} \\
  {\rm where, \ {\boldsymbol X} } = \Big[ \ln(n_e)-\mu_{ne} \hspace{.25cm},\hspace{.25cm} \ln(T)-\mu_{T} \Big] \nonumber
\end{eqnarray}
and ${\boldsymbol {\mathcal C}}$ is the covariance matrix,
\begin{equation}
 {\boldsymbol {\mathcal C}} = \left[ \begin{array}{cc} \sigma_{ne}^2 & \sigma_{ne}\sigma_{T}\xi \\ \sigma_{ne}\sigma_{T}\xi & \sigma_{T}^2 \end{array} \right]. \nonumber
\end{equation}

 Such a distribution is characterised by the means of the lognormal distribution,
 $\mu_{ne} = \langle \ln(n_e/ {\rm 1 \times 10^{-3} \  cm^{-3}}) \rangle$ and $\mu_{T} = \langle \ln(T/1 \ {\rm keV}) \rangle$ \alert{(or $T_{\rm median} = \exp(\mu_T)~{\rm keV}$)},
 the variances, $\sigma_{ne}^2$ and $\sigma_T^2$, along with the coefficient of correlation, $\xi$, between $\ln(n_e)$ and $\ln(T)$. Of these, one may always
 set the density parameter, $\mu_{ne}$ to an arbitrary value since we are only interested in computing the ratio $P_X$/$P_{SZ}$; henceforth we shall not
 mention the value of this parameter.

\begin{figure}
 \centering
  \hspace*{-4.5mm}
  \includegraphics[width=0.54\textwidth]{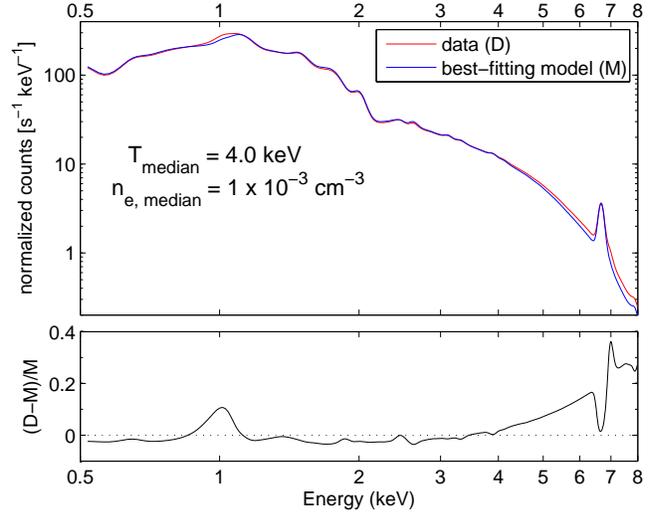}
  \caption{Single temperature fit to a composite spectrum, for the Chandra response.
{\it Upper panel:} The red curve shows the composite spectrum generated from a bivariate lognormal distribution of gas components  with $\sigma_T=0.45$, $\sigma_{ne}=0.55$
 and correlation $\xi=-0.55$, while the blue curve is the best-fitting single temperature MEKAL spectrum. {\it Lower panel:} Difference between the data and model.
 For this spectrum the best-fitting values of density and temperature are 1.38$\times$10$^{-3}$ cm$^{-3}$
 and 3.91 keV, giving biases of +16.8\% and -23.3\% w.r.t. $n_{\rm ref}=1.16 \times 10^{-3}~{\rm cm}^{-3}$ and $T_{\rm ref}=3.91~{\rm keV}$
  respectively (see equation \ref{eqn:nref}). Using equation \ref{eqn:press_bias}, the bias in pressures turns out to be -10.4\%.
 A metallicity value of 0.35 was assumed relative to the solar abundance \citep{solar}.
}
  \label{fig:xspec}
\end{figure}

\subsection{Method: Mock X-ray spectrum}
\label{sec:xspecmethod}
We use the MEKAL code in XSPEC
 \citep{1985A&AS...62..197M,1986A&AS...65..511M,Kaastra,1995ApJ...438L.115L} to generate the X-ray model spectra from a hot diffuse gas, including the line emissions
 from the astrophysically abundant elements, for each of the components defined by the values of $n_e$ and $T$. The choice of the MEKAL model was motivated by
its ability to calculate anew the spectrum at any given temperature, rather than interpolate from tabulated values as in the APEC model\footnote{We find that
 the APEC model \citep{APEC} produces small spurious jumps in the bias caused by the interpolation of the tabulated spectra. These jumps are absent in the MEKAL
 model if the emission spectra are calculated anew for each model.}. These spectra are then added together to produce a composite spectrum of X-ray emission from
 an ICM having a distribution of temperatures and densities. This composite spectrum is next fitted by a single component MEKAL model (in the energy range 0.5 - 8.0 keV)
 to obtain the best-fitting values $n_{\rm e, fit}$ and $T_{\rm fit}$ using the Chandra instrumental response file. The \alert{X-ray electron pressure, $P_X=n_{e \rm, fit}T_{\rm fit}$},
 for the ICM is then computed from these best-fitting values of density and temperature\footnote{Note that when fitting a MEKAL model to the composite spectrum
 we allow only the temperature and normalisation to be free parameters, while the redshift, metal abundances, and the Hydrogen column density are kept fixed.}.
 
 The bias in the gas density is defined with respect to the volume-weighted density and the bias in temperature is defined with respect to the mass weighted temperature,
\begin{equation}
\label{eqn:nref}
\alert{ n_{\rm ref} = \frac{\int n_e dV}{\int dV} ~~~~ ; ~~~~ T_{\rm ref} = \frac{\int T n_e dV}{\int n_e dV}.}
\end{equation}
With these definitions the SZ pressure is, \alert{$P_{SZ} = P_{\rm mean} = n_{\rm ref} T_{\rm ref}$} \alert{and the bias in pressure may be written as,}
\begin{equation}
 \label{eqn:press_bias}
 \alert{ b_P \equiv \frac{P_X}{P_{SZ}} - 1= \frac{n_{e \rm, fit}}{n_{e, \rm ref}} \frac{T_{\rm fit}}{T_{\rm ref}} - 1}.
\end{equation}
 The top panel of Fig. \ref{fig:xspec} shows the mock X-ray
 spectrum generated from a hot diffuse gas that has a bivariate lognormal distribution (red line) with a median temperature, $T_{\rm median} = 4$ keV and
 median density, $n_{e, {\rm median}} = 1 \times 10^{-3}$ cm$^{-3}$. The blue line shows the best-fitting spectrum specified by a single
 value of temperature and density. Using equation \ref{eqn:press_bias} we compute the pressure bias, $b_P = -10.4\%$.

\begin{figure*}
 \centering
 \hspace*{-5mm}
 \psfrag{cc}[\vspace{0mm}][][1.4]{$\mathbf{T_{\rm median} = 0.5 ~{\rm keV}}$}
\psfrag{dd}[\vspace{0mm}][][1.4]{$\mathbf{T_{\rm median} = 1.0 ~{\rm keV}}$}
\psfrag{ee}[\vspace{0mm}][][1.4]{$\mathbf{T_{\rm median} \geq 3.0 ~{\rm keV}}$}
 \psfrag{bb}[\vspace{-3mm}][][1.4]{$\mathbf{ Z/Z_{\odot} = 0.2}$}
 \psfrag{aa}[\vspace{-2mm}][][1.4]{$\mathbf{ Z/Z_{\odot} = 0.5}$}
 \includegraphics[width=1.05\textwidth]{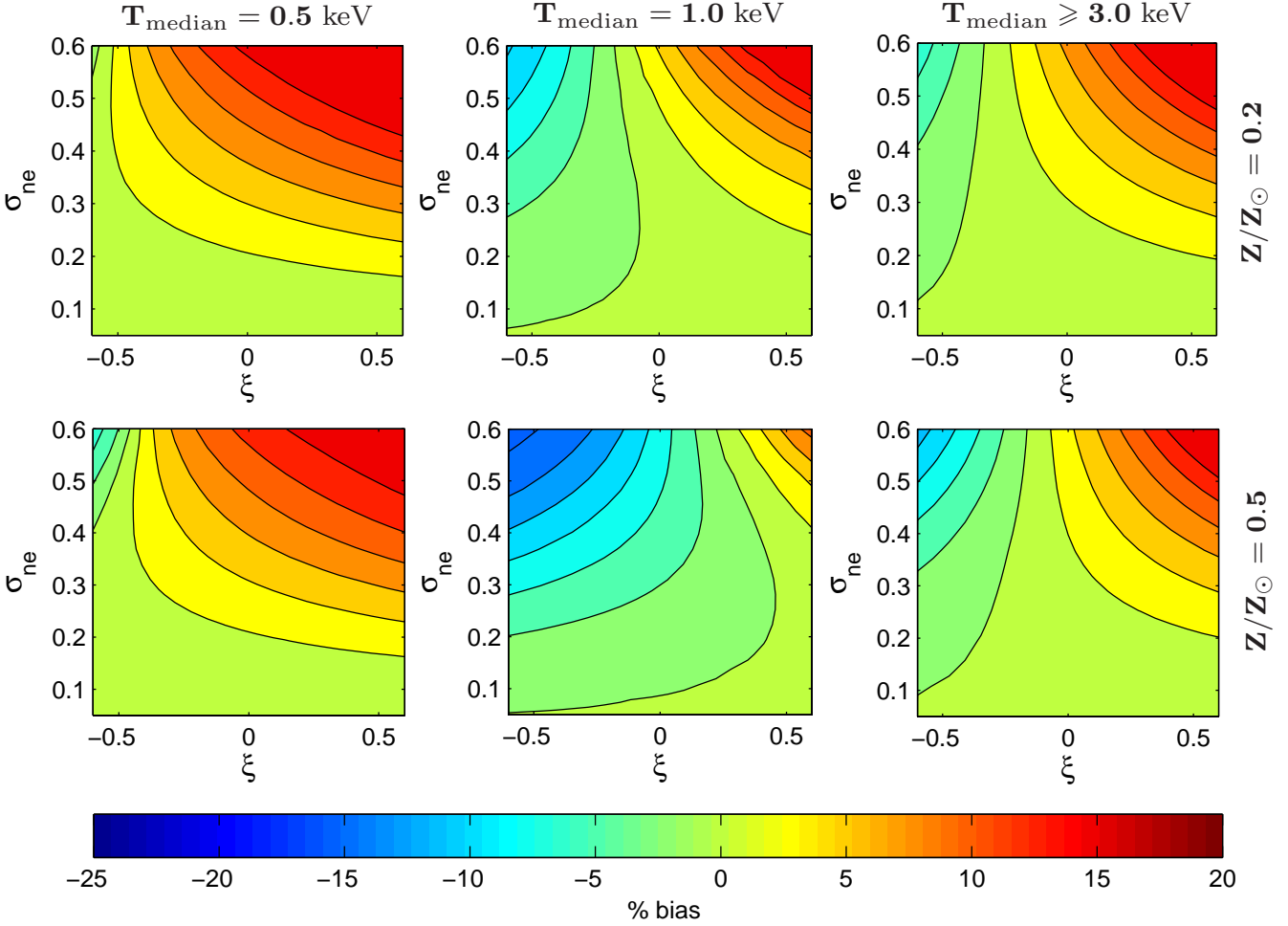}
\caption{Contour map predicting the bias, \mbox{$b_P \equiv P_{X}/P_{SZ} - 1$}, for an idealised hot gas described by a bivariate lognormal distribution
of density and temperature fluctuations for various values of median temperatures, $T_{\rm median}$ (in columns) and assumed metallicities (in rows). The bias
values are indicated on the colourbar (in percentage) as a function of two parameters $\sigma_{ne}$ and $\xi$, \alert{using the best fitting relation
from Fig. \ref{fig:sigma_plot_col}, $\sigma_{T} = 0.73\sigma_{T} - 0.02$}. The green contours indicate the zero bias, while
 adjacent contours are in increments of 2.5\%. At higher values of $T_{\rm median}$, the dependence of bias on $\xi$ and $\sigma_{ne}$ is very
 similar to that indicated by $T_{\rm median}=3$ keV. The bias has been computed by fitting a single
 temperature spectrum in the energy band 0.5--8.0 keV and using the Chandra response function.}
\label{fig:contour}
\end{figure*}

\subsection{Results: Bias from a lognormal distribution of gas density and temperature fluctuations}
As a preliminary check we verified that our scheme reproduces
 the simple estimate of the ratio \alert{\mbox{$P_X/P_{SZ} = n_{e, \rm fit}/n_{e, \rm ref} = \sqrt{C}$}} when the temperature
 fluctuations are set to zero, while retaining only the density perturbations \alerta{(}see also equation \ref{eqn:estimate}\alerta{)}.
 \alert{This is done by generating a mock spectrum from a lognormal distribution of density fluctuations, convolving it with
 the instrument response (Chandra response in our case) and then fitting it to a single temperature model, as explained before
 in section \ref{sec:xspecmethod}, to compute the bias using equation \ref{eqn:press_bias}.} We then go on to
 systematically explore the dependence of bias in an \alert{idealised} hot gas described by a \alert{bivariate lognormal distribution of} density
 and temperature fluctuations \alerta{(}see equation \ref{eqn:lognormal}\alerta{)}.
 For this we compute the bias \alert{(as described in section \ref{sec:xspecmethod}) as a function of the parameters
 $\{ \sigma_{T}, \sigma_{ne}, \xi, T_{\rm median} \}$ for} values in the range -- $0.01 \leq \sigma_{T},~ \sigma_{ne} \leq 0.6$,  and
 $-0.6 \leq \xi \leq 0.6$ at various temperatures, \alert{$T_{\rm median} \geq 0.5~{\rm keV}$}, for an assumed value of
 metallicity. We find that the bias is most sensitive to the value of $\xi$, the coefficient of correlation between density and
 temperature fluctuations (see equation \ref{eqn:lognormal}). Keeping the other parameters fixed, the bias always increases almost
 linearly with $\xi$ for $|\xi| < 0.55$. The rate of increase however depends on the widths of the distribution of temperature and
 density fluctuations, i.e. $\sigma_T$ and $\sigma_{ne}$. The simple linear dependence of the bias on $\xi$ implies the possibility
 of probing the nature of perturbations in the ICM: $\xi < 0$ would indicate the relative importance of isobaric or entropy\footnote{Here
 entropy is defined as $S \equiv k_B T/n_e^{2/3}$, where $n_e$ and $T$ are electron density and temperature of the ICM respectively.}
 perturbations, while $\xi > 0$ would imply that adiabatic or pressure perturbations (caused by sound waves or weak shocks) are dominant.

We now try to describe the dependence of the bias in pressure in terms of the parameters of the bivariate lognormal distribution using a fitting form. For
 values of the median gas temperature, $T_{\rm median} \geq 3$ keV, the bias may be parametrised \alerta{using three free parameters as follows},
\begin{equation}
\label{eqn:fitformeqn}
\alert{  b_P = \frac{1}{2} \sigma_{ne}^2 - p_1 \sigma_T^2 + p_2\ln(1+p_3\sigma_T) \sigma_{ne} \xi.}
\end{equation}
\alerta{The best-fitting} values of the parameters are indicated in Table \ref{tab:fittingform} for a range of assumed metallicites.
This fitting formula is accurate (in absolute terms) to within 0.01 for $\sigma_{ne/T} \leq 0.3$ with $|\xi| \leq 0.6$ and within 0.05 for $\sigma_{ne/T} \leq 0.6$ with
 the same range for $\xi$. \alerta{We verified that these conditions hold for the ICM within $R_{500}$, in simulated clusters (see Section \ref{sec:results_bias}
 and the discussion on Fig. \ref{fig:rad_profile})}.
\begin{table}
 \centering
 \caption{Values of the best-fit parameters of the fitting form in equation \ref{eqn:fitformeqn} to predict the bias, $b_P \equiv P_X/P_{SZ}-1$, in X-ray vs SZ pressure,
 valid at temperatures, $T_{\rm median} \geq 3$ keV.}
 \begin{tabular}{@{}lccc@{}}
 \hline
 $Z/Z_{\odot}$ & $p_1$ & $p_2$ & $p_3$\\
\hline
0.20 & 0.69  & 0.75 & 1.48\\
0.35& 0.75  & 0.83 & 1.31\\
0.50 & 0.80  & 0.99 & 1.08\\
\hline
\label{tab:fittingform}
\end{tabular}
\end{table}
For median temperatures greater than 3 keV, the X-ray emissivity in the 0.5--8 keV band is only a weak function of temperature;
 at lower temperatures however, there is a significant dependence due to relative importance of discrete spectral line emission, causing the measured bias
 to depart significantly from this parametrisation. We find that a decrease in the median temperature below $\sim 1$ keV causes the bias to change steeply
 in a non-trivial manner, which is difficult to parametrise through a simple form.

In general the bias in pressure {for an assumed value of metal abundances} depends on 4 parameters -- $\sigma_T$, $\sigma_{ne}$, $\xi$ and $T_{\rm median}$.
 However, we may simplify this by eliminating a parameter using an approximate relation, \alert{$\sigma_{T} = 0.73\sigma_{T} - 0.02$}, that is seen in
 hydrodynamical simulations of clusters (see Fig. \ref{fig:sigma_plot_col}).
With this simplification, in Fig. \ref{fig:contour} we plot the bias contours for various values of median temperatures as a function
 of $\xi$ and $\sigma_{ne}$ assuming two values of gas metallicities, $Z/Z_{\odot} = $ 0.2 and 0.5; the former being the typical value beyond $0.2~R_{180}$,
 while the latter is indicative of the ICM metallicities at the centre of the cluster \citep[e.g.][]{2008A&A...487..461L}. We see that for ICM temperatures,
 $T_{\rm median} > 3$ keV, the bias is small (from $-10\%$ to $+5\%$) for most of the parameter space in the ($\xi$, $\sigma_{ne}$) plane.
 Only for higher values of both the fluctuations and the negative correlation $\xi$, the bias goes beyond $-15\%$. For lower temperatures
 (1 keV $ < T_{\rm median} <$ 3 keV), this bias can be between $-25\%$ to $+15\%$, depending on the metal abundance; while for even lower temperatures
 the bias turns positive, mostly lying at the $0 - 10\%$ level. Note that especially for low temperatures ($T_{\rm median} < 3$ keV) the
 value of bias is more sensitive to the assumed metal abundance; the effect of increasing metallicity is to reduce the bias \alerta{when positive},
 or to move it towards more negative values \alerta{otherwise}.

\section{Simulations and sample of Galaxy Clusters}
 \label{sec:simulations}
\begin{table}
 \centering
 \caption{Properties of simulated clusters in the sample of \citet{Nagai2007b, Nagai2007a} at $z=0$. Two runs were performed with different physics,
CSF run with cooling+star formation and NR run without radiative cooling and star formation. The last column indicates
 if the cluster is visually identified as relaxed (R) or unrelaxed (U).}
 \begin{tabular}{@{}rccc@{}}
 \hline
 Cluster ID & R$_{500}$  & \alert{$M_{500}^{\rm tot}$} & Relaxed (R) or \\
  &  ($h^{-1}$ Mpc) & \alert{$(10^{14}~h^{-1}~{\rm M}_{\odot})$} & Unrelaxed (U) \\
\hline
  & CSF / NR & CSF / NR & CSF / NR \\
\hline
CL101 &   1.16 / 1.14  & 9.08 / 8.62 & U / U \\
CL102 &   0.98 / 0.95  & 5.45 / 4.63 & U / U \\
CL103 &   0.99 / 0.99  & 5.71 / 5.71 & U / U \\
CL104 &   0.97 / 0.97  & 5.39 / 5.31 & R / R \\
CL105 &   0.94 / 0.92  & 4.86 / 4.50 & U / U \\
CL106 &   0.84 / 0.84  & 3.47 / 3.40 & U / U \\
CL107 &   0.76 / 0.78  & 2.57 / 2.74 & U / U \\
CL3     &   0.71 / 0.70  & 2.09 / 1.98 & R / R \\
CL5     &    0.61 / 0.61 & 1.31 / 1.34 &  R / U \\
CL6     &    0.66 / 0.61 & 1.68 / 1.32 &  U / R \\
CL7     &    0.62 / 0.60 & 1.41 / 1.25 &  R / R \\
CL9     &    0.52 / 0.51 & 0.823 / 0.775 &  U / U \\
CL10   &    0.49 / 0.47  & 0.672 / 0.621 &  R / R \\
CL11   &    0.54 / 0.44  & 0.899 / 0.482 &  U / R \\
CL14   &    0.51 / 0.48  & 0.769 / 0.652 &  R / R \\
CL24   &    0.39 / 0.39  & 0.347 / 0.347 &  U / U \\
\hline
\label{tab:sample}
\end{tabular}
\end{table}

We use a sample of 16 simulated clusters at $z=0$ taken from \citet{Nagai2007b,Nagai2007a}. The simulations were done using the Adaptive Refinement Tree (ART)
 N-body+gas-dynamics code \citep{1997ApJS..111...73K, 2002ApJ...571..563K} assuming a flat $\Lambda$CDM cosmology with the following values for the
 cosmological parameters: $\Omega_m=0.3, \Omega_b=0.04286, h=0.7$ and $\sigma_8=0.9$. Two types of simulations were used with different physics involved,
 but with the same initial conditions: \mbox{(i) non-radiative} (NR) runs without any radiative cooling and \mbox{(ii) cooling} plus star formation (CSF) runs,
 which included metallicity-dependent radiative cooling, star formation, supernova feedback and UV background. The clusters
 in the sample were selected randomly to uniformly sample the mass range,
 $ 7\times 10^{13}~h^{-1}~{\rm M_\odot} < M_{500} < 2\times 10^{15}~h^{-1}~{\rm M_\odot}$.
 These clusters were selected from low-resolution simulations and resimulated at a higher resolution. The division of the sample into relaxed
 and unrelaxed sub-samples (see Table \ref{tab:sample}) was done in \citet{Nagai2007b} by visually examining the morphology of mock X-ray images
 in three different projections. \alert{This was done by producing mock images of the simulated clusters using the Chandra response function with a
 100 ks exposure. From these images the `relaxed' clusters were identified as those with a regular X-ray morphology and without any significant
 deviations from elliptical symmetry. On the other hand, clusters were classified as `unrelaxed' when departures from elliptical symmetry,
 filamentary X-ray structures, or large isophotal centroid shifts were seen in the produced images. This procedure is similar to that used by
 observers. Z12 made another classification based on the widths of the density distribution, $\sigma_{ne}$, at $r=R_{500}$.
This classification agrees very well with that done in \citet{Nagai2007b}, and the difference in $\sigma_{ne}(R_{500})$ between relaxed and
 unrelaxed clusters was at the level of $\backsimeq 2.5 \sigma$. There also exist other schemes of classification, for e.g. using the two
 dimensional multipole expansion of the projected gravitational potential \citep{Buote} or computing the centroid shifts \citep{Poole};
 however we did not compare our classification with these.}

\alert{The ART code uses a non-uniform and dynamic mesh adjusting to the evolving particle distribution, thus allowing for a large dynamic range.
To simplify our analysis, the code output (gas densities, temperatures, velocities, and metallicities) was interpolated onto a uniform grid with grid cells of size comparable to the effective resolution of the cluster simulations.} Following Z12, we randomly sample $4\times 10^7$ data points within the
sphere of radius of $5~h^{-1}~{\rm Mpc}$ with a weight $\propto 1/r^2$,
where $r$ is the distance from the cluster centre. This sampling is uniform in azimuthal and polar directions and provides equal number of
points per spherical shell of a given thickness. \alert{Such a scheme is useful for dealing with averaged quantities computed over spherical shells
as discussed further in the following sections.}

\section{The bias from simulated Galaxy Clusters}
\label{sec:results}
\alerta{In this section we shall compute the bias between `measured' pressure profiles from X-ray and SZ observations using the sample of simulated clusters
 introduced in the previous section. We wish to emphasise that the term `measured' (here and henceforth) refers to the value of pressure as would be inferred
 from observations in X-ray/SZ through a deprojection analysis of real data. Note that this may well be different from the actual values of
thermal pressure of the ICM.}

Each cluster is divided into 32 concentric shells (spaced in equal logarithmic
 intervals) between the radii $0.1 <r/R_{500} < 2.3$, in which we compute the histogram of density and temperature values. This is then used to
 generate fake spectra assuming\footnote{Strictly speaking, assuming a fixed metallicity value for different shells is
reasonable only within $\sim 0.5 ~R_{500}$ where $T_{\rm median} > 1~ {\rm keV}$ for most of the simulated clusters, and the bias
 depends only weakly on the assumed metal abundance. For $T_{\rm median} \lesssim 1$ keV, we find that our results on bias are more influenced
by metallicity; when higher accuracy is required, especially so for lower temperatures, one has to \alert{do} the direct analysis on a case by case
 basis.} a metallicity of $Z/Z_{\odot}=0.35$, as described before in section \ref{sec:xspecmethod}.
 The fake spectra are then analysed using the Chandra response matrix to give the best-fitting values of density, $n_{e, {\rm fit}}$ and temperature,
 $T_{\rm fit}$ for each shell. The `measured' X-ray pressure is then simply $P_X=n_{e, {\rm fit}}T_{\rm fit}$. On the other hand the SZ pressure for
 each shell is obtained by computing the mean thermal pressure in the shell $P_{SZ}=\sum_{i} n_{e,i} T_i w_i/\sum_{i} w_i$, where $w_i$ is the weight,
 i.e. the fractional volume occupied by the ICM specified by the values $n_{e,i}$ and $T_i$ for density and temperature respectively. These are used
 now to compute the bias, $b_P(r)=P_X(r)/P_{SZ}-1$ between the `measured' X-ray and SZ pressures, in each shell.

 \begin{figure*}
  \centering
  \hspace*{-5mm}
  \subfloat[$\sigma_{ne}(r)$]{\includegraphics[width=.54\textwidth]{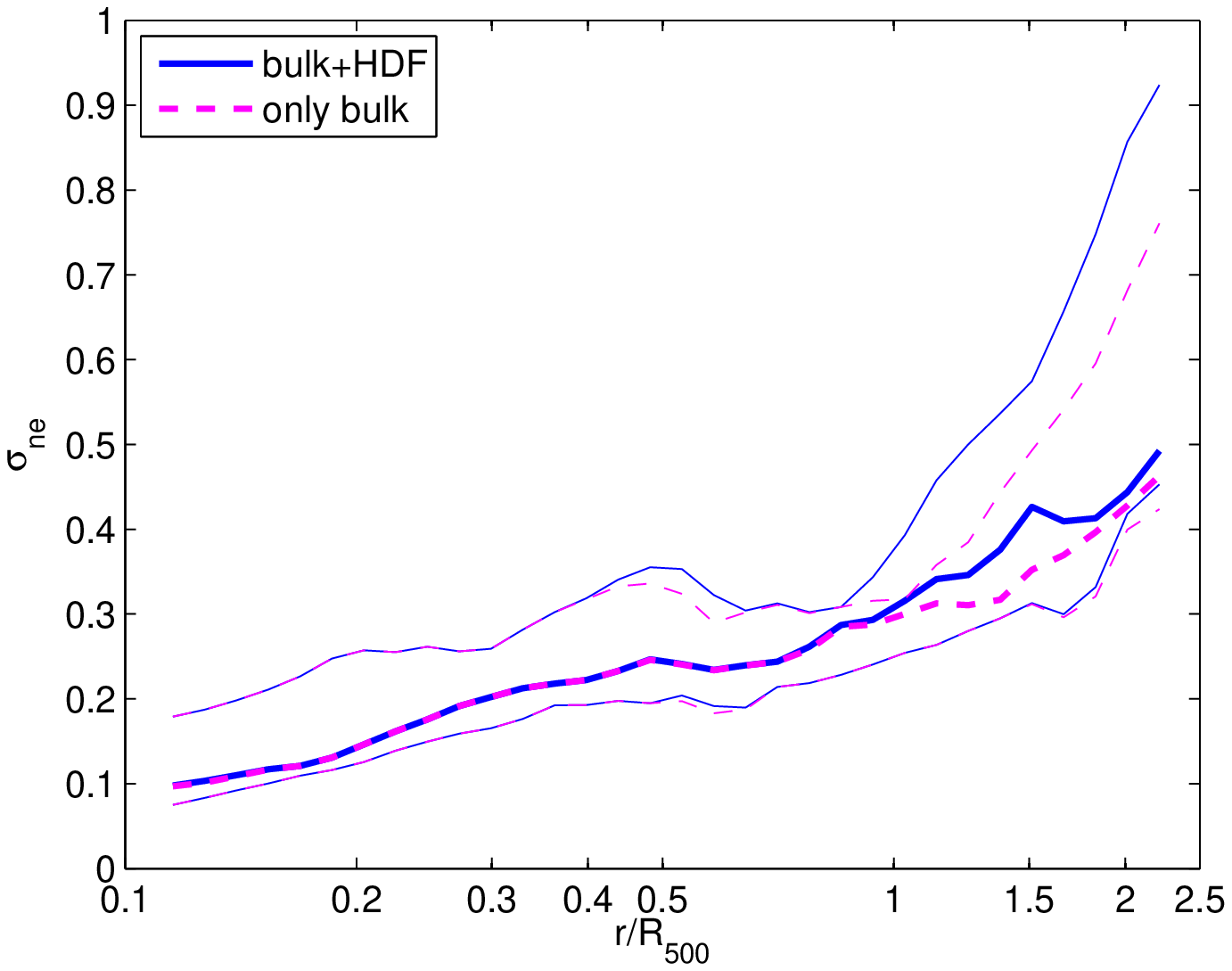}}
  \hspace*{-2mm}
  \subfloat[$\xi(r)$]{\psfrag{abcd}{$\xi {\big(} \ln(n_e),~\ln(T) {\big)}$}\includegraphics[width=.54\textwidth]{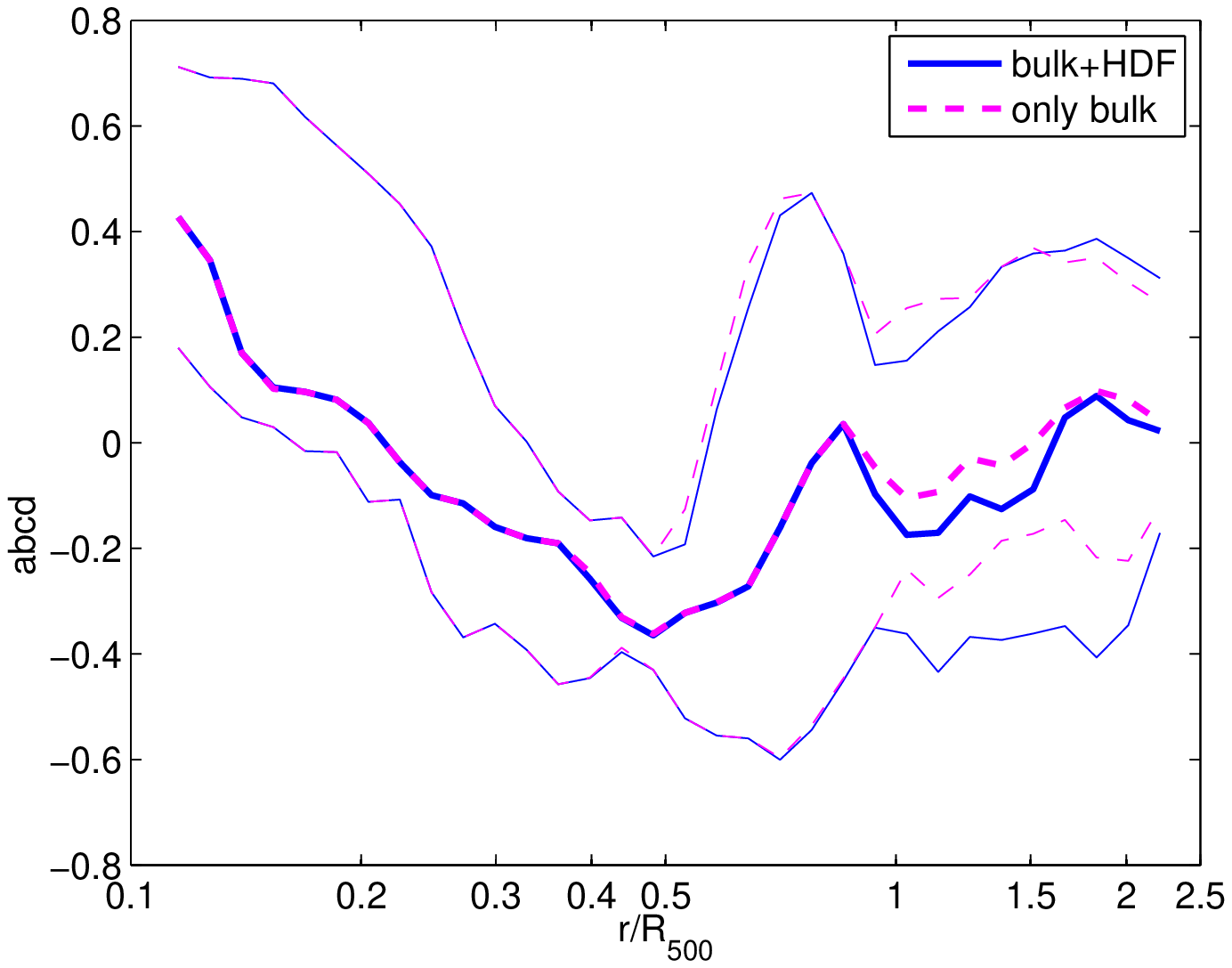}}
  \caption{Radial profiles showing the rms of fluctuations in density, $\sigma_{ne}$ (characterised by a lognormal
 distribution), for a sample of relaxed clusters in the NR simulations. The thick lines indicate the sample median while the thin lines
denote the sample outliers. The solid and dashed lines correspond to the ICM with and without high density fluctuations (HDF), respectively.}
  \label{fig:rad_profile_others}
\end{figure*}

\subsection{Properties of the ICM fluctuations}
\label{sec:prop}
Z12 used the same set of simulations, as described in section~\ref{sec:simulations}, to study the properties of the inhomogeneities in the ICM.
 They developed a simple and robust method to separate the HDF from the nearly hydrostatic bulk component. This was implemented by
 imposing a cut within each shell to exclude all particles having densities, $n_e$, such that $\log(n_e) > \log(n_{e, {\rm median}}) + 3.5 \sigma_{ne}$.
Henceforth we shall use the term `HDF' to exclusively denote the particles in the simulation lying above this cut, and the term `bulk' to describe
 the particles lying within this cut.

\begin{figure}
  \vspace{-2mm}
  \centering
  \hspace*{-5mm}
  \includegraphics[width=0.54\textwidth]{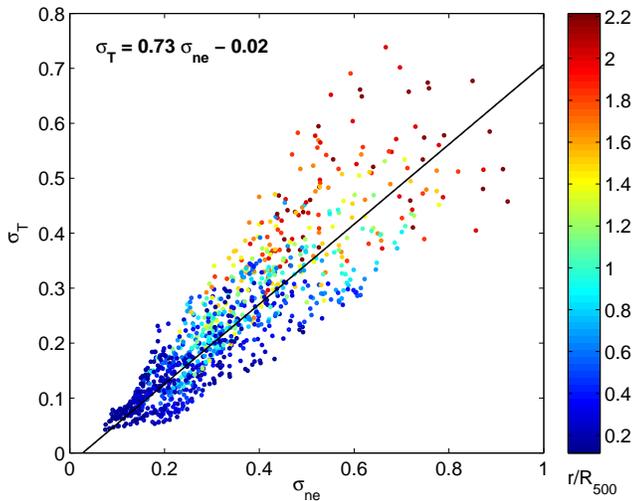}
  \caption{The coloured dots indicate the values of fluctuations in density, $\sigma_{ne}$, and temperature, $\sigma_T$, in
 concentric shells from the combined sample of clusters in both the NR and CSF simulations. The colours (mapped to colourbar on right) indicate the radial
 distance of the shell from the cluster centre. The black line indicates \alert{the best fitting linear relation, $\sigma_{T} = 0.73\sigma_{ne}-0.02$.
 For this relation the scatter in $\sigma_T$ for a given value of $\sigma_{ne}$ is 0.07.}}
  \label{fig:sigma_plot_col}
\end{figure}

Here we review some of the results presented in Z12. Fig. ~\ref{fig:rad_profile_others} shows the
 radial profiles indicating the properties of ICM fluctuations for the sample of relaxed clusters in the NR simulations.
 In Fig. ~\ref{fig:rad_profile_others} (a)
 the solid lines denote the median values (over the sample of relaxed clusters in the NR run) of the parameters obtained by fitting a bivariate lognormal
 distribution (see equation~\ref{eqn:lognormal}) within each shell. The dashed lines are the same quantities computed {\em after} the removal
 of the HDF components. The thin lines indicate the maximum and minimum outliers of the sample that shows the scatter across the sample.
 Fig. ~\ref{fig:sigma_plot_col} shows the relation between the parameters $\sigma_{ne}$ and $\sigma_T$ in
 each shell; the straight line is the best-fitting linear relation, \alert{$\sigma_{T} = 0.73\sigma_{ne}-0.02$}, which is \alert{also the} assumption
 used in producing the contours in Fig. ~\ref{fig:contour}. \alert{The data from NR and CSF simulations indicate an intrinsic scatter of 0.07
in $\sigma_T$ for a given value of $\sigma_{ne}$ in this relation.}
 Fig.~\ref{fig:rad_profile_others} (a) and (b) show that the values of the parameters, $\sigma_{ne}$ and $\xi$, are fairly robust to the exclusion
 of the HDF components from the bulk.

In both the CSF and NR runs, the unrelaxed clusters have not only higher values of temperature and density fluctuations, but also show significantly larger scatter
from cluster to cluster. Both relaxed and unrelaxed clusters show a steady increase in the fluctuation amplitude with radius, which steepens in the
 outskirts ($r > R_{500}$). As shown in Fig. ~\ref{fig:rad_profile_others} (a), the fluctuations in gas density, $\sigma_{ne}$, are large beyond
 $R_{500}$, and these fluctuations are reduced if the HDF components are removed. Fig. ~\ref{fig:rad_profile_others} (b) shows that the correlation between temperature
 and density fluctuations, $\xi$, is mostly negative, but increases towards zero at larger radii. The increase in $\xi$ on removal of high density clumps is in line
 with our expectations, as the clumpy regions contribute to an anti-correlation (or $\xi < 0 $) between density and temperature fluctuations.
 
\subsection{Removing the contribution from the brightest clumpy regions in the cluster outskirts}
\label{sec:clumpy}
We expect the `measured' quantities related to the ICM (e.g., gas pressure, density and temperature) to be less biased when the HDF components
are removed from the simulation data. Clearly the removal of HDF according to the criterion described in the previous section is only possible
 in simulations. In real observations, the complete removal of HDF seems unlikely, due to the presence of the X-ray background noise and limited photon
 statistics, especially relevant at the cluster outskirts or for the high redshift clusters. Here we shall attempt to check this by dealing directly
 with the 2-dimensional projected X-ray images obtained from the simulated clusters, only in the limit of infinite photon statistics (i.e., without any Poisson
 noise) and in the absence of instrumental background. However, we shall argue in section \ref{sec:results_bias} that even a partial removal of the clumps,
 especially the brightest ones (as permitted by the photon statistics), helps to reduce the bias to a large extent.
\begin{figure*}
 \centering
  \includegraphics[width=1\textwidth]{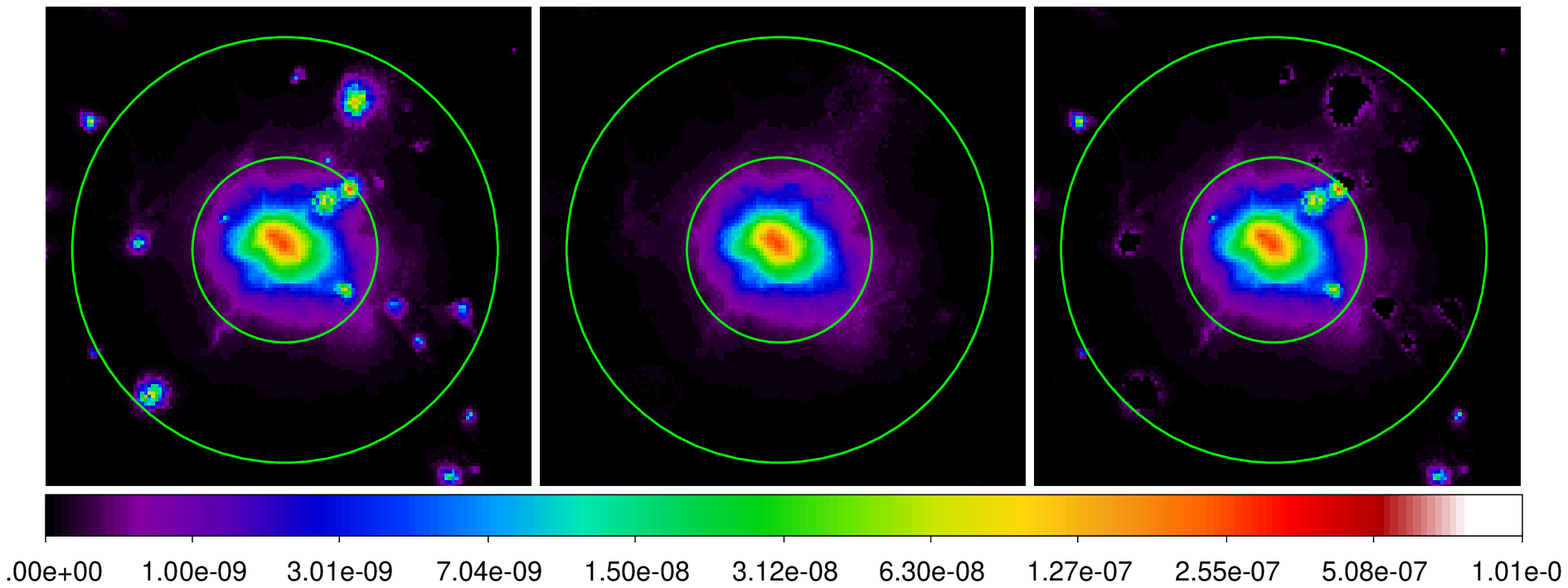}
  \caption{Exclusion of the bright clumps from the projected \alert{soft band (0.5--2.0 keV)} \mbox{X-ray} image of the simulated cluster CL101 in the NR run.
  \alert{The images from left to right are: \mbox{(i) Original} image; \mbox{(ii) Image} showing only the bulk component (i.e. with 3-D cut); \mbox{(iii) Image}
 after masking of the brightest regions (i.e. with 2-D cut specified by the value $f_{\rm cut} = 2.5$).} The 2-D mask is applied in the region
 $R_{500} < r < 2.3 ~R_{500}$ whose boundaries are indicated by the green circles, \alerta{while the 3-D mask is applied everywhere.}}
  \label{fig:mask}
\end{figure*}

\begin{figure*}
  \centering
  \hspace*{-3.5mm}
  \includegraphics[width=1.03\textwidth]{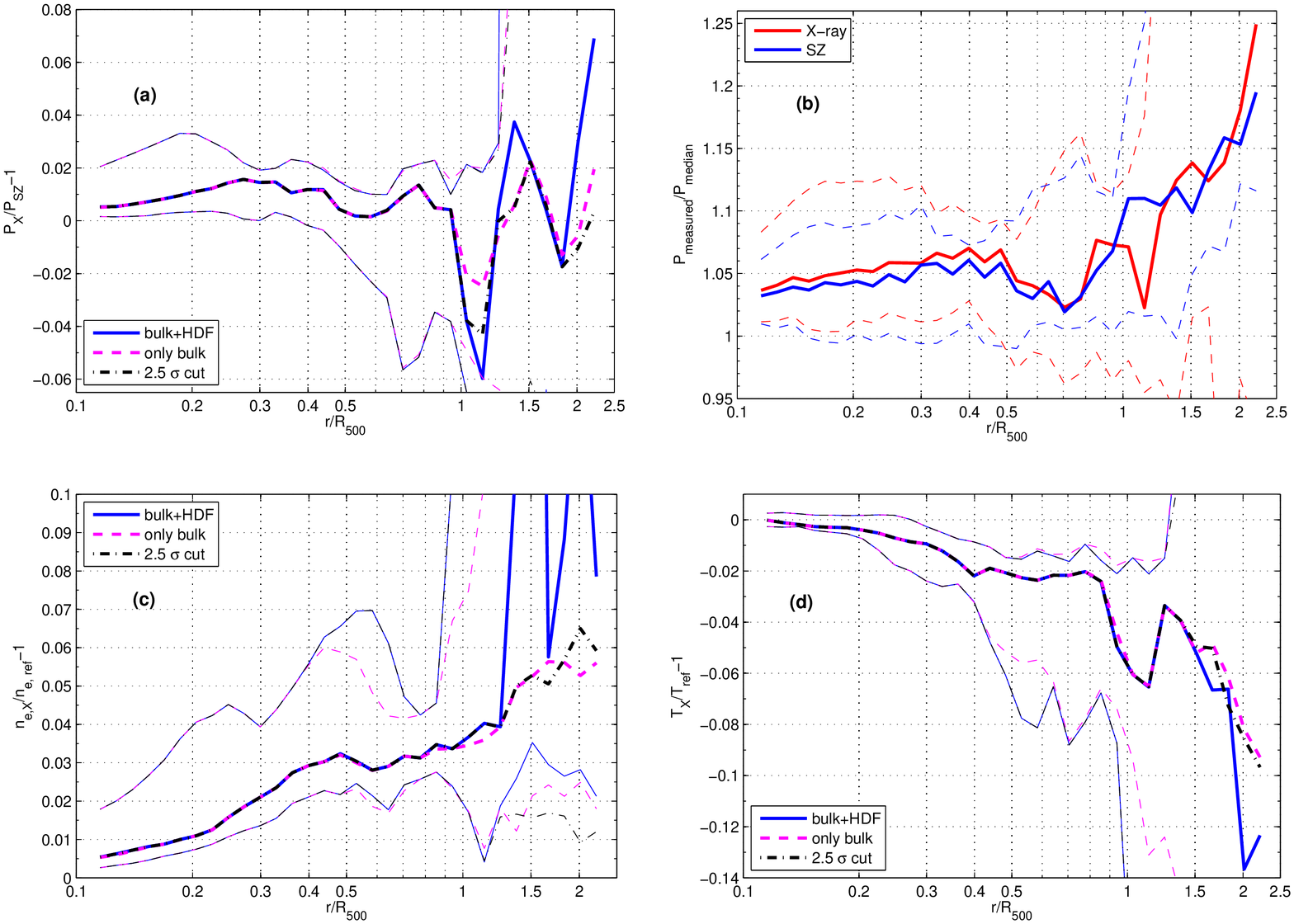}
  \caption{Radial profiles showing various biases computed in concentric radial shells of a sample of relaxed clusters in the non-radiative simulations.
 In all the figures, the thick lines indicate the median profile while the thin lines are the most extreme outliers in the sample.\\
(a) \alerta{Pressure bias}, $b_P \equiv P_X/P_{SZ}-1$, for the `measured' X-ray ($P_X$) and SZ ($P_{SZ}$) pressures. \\
(b) The ratio of the `measured' X-ray/SZ pressures to the median pressure shown in thick red/blue lines for `bulk+HDF'. \\
(c) \alerta{Bias in the} `measured' density from X-ray observations with respect to $n_{\rm ref}$ \alert{(see equation \ref{eqn:nref})}. \\
(d) \alerta{Bias in the} `measured' temperature with respect to the mass weighted temperature,
 $T_{\rm ref}$ \alert{(see equation \ref{eqn:nref}). The `measured' temperature is determined simultaneously with the best-fitting $n_e$.}\\
}
\label{fig:rad_profile}
\end{figure*}

For each of the clusters, we divide the X-ray images into six radially equal annular regions starting from $r=R_{500}$ as the innermost
 ring and up to $r=2.3 ~R_{500}$. \alert{We choose $R_{500}$ as the inner radius for the purpose of removing clumps since we find that the bias outliers
 become large only beyond this radius, suggesting that the effect of clumps is dominant in the bias profiles only in the outskirts ($r > R_{500}$) of clusters
 \citep[see also][and Fig. \ref{fig:rad_profile}]{1999ApJ...520L..21M, 2011ApJ...731L..10N, Battaglia, 2012arXiv1211.1695V}.}
Using the pixels
 lying within each annular region we then create histograms of the X-ray brightness (in the soft band
 from 0.5-2.0 keV using the Chandra response) and mask all pixels with brightness, $B$, such that \alert{\mbox{$\log(B) > \log(B_{\rm median}) + f_{\rm cut} \sigma_{B}$}},
 to zero values, \alert{where $\sigma_B$ is the standard deviation of the logarithm of pixel brightness in each annulus}. Using three different
 values of $f_{\rm cut}$ -- 2, 2.5 and 3, we then eliminate all the particles from the simulation data which
 lie exactly along the line of sight of the masked pixels, some of which are responsible for producing the very bright regions in the outskirts of
 the X-ray image. We shall henceforth refer to this method of removing the high-density contributions from the X-ray images as the 2-D cut in
 contrast with the 3-D cut performed directly on the radial shells and as outlined in section \ref{sec:prop} and detailed in section 4 of Z12.

We note that the 2-D cuts do not remove all of the clumpy contributions from the 3-dimensional radial shells, partly because clumps
 lying along the line of sight through the bright central region of the projected image of the cluster would not show up with high enough contrast
 in the brightness images, and would remain in the filtered (i.e. after a 2-D cut) simulation data. \alert{For a fair comparison of the results from
2-D cut and 3-D cut we attempt to correct this by applying a projection in another plane and repeating the masking procedure in this new plane.}
\alerta{Fig. \ref{fig:mask} shows a comparison of projected X-ray images of the simulated cluster CL101 under 2-D (using $f_{\rm cut} = 2.5$) and 3-D cuts.
On the left is the original image produced without any cuts, while the central and the rightmost images are after application of 3-D and 2-D cuts respectively.
 Note that the 3-D cut is applied everywhere while the 2-D cut is only applied within $R_{500} < r < 2.3 ~R_{500}$
 (indicated by the annulus which is bounded by green circles). We see that the bright clumps in this annulus, visible prominently on the leftmost image
 are absent after applying either the 3-D or 2-D cuts. The fact that this annulus appears very similar after 2-D and 3-D cuts
 indicates that both of these cuts remove nearly the same points from the 3-D shells.}

\alerta{We now try to examine if a similar 2-D cut (or masking) is possible in real observations, where the ability to mask the clumps would depend on the
 signal to noise ratio at $r \gtrsim R_{500}$. To be able to identify and mask such regions it is necessary that the intrinsic fluctuations of surface
 brightness should dominate over the Poisson fluctuations of the photon counts.} We now estimate the minimum number of counts required in order to be able
 to mask such clumps, provided that they are resolved by the instrument. Let $N$ be the total number of counts, $n_{\rm reg}$ be the number of independent regions (or
 pixels used for masking); then, we have $n=N/n_{\rm reg}$ as the number of counts per region. To reliably remove the region with the surface brightness $X$ times
 the mean value (i.e. $Xn$ counts) one needs (case with no background) as a conservative estimate: $(X-1)n>5\sqrt{n}$ for $n>>1$, or $(X-1)n>5$ for
 $n<<1$. For the case shown in Fig. \ref{fig:mask} we use 63 by 63 kpc pixels, yielding $N_{\rm reg}= 9048$ in the annulus from $R_{500}$ to $2.3~R_{500}$.
 Assuming 100,000 counts for the whole image, and 7.0\% flux coming from this annulus, the typical number of counts per pixel, $N/n_{\rm reg}=0.78$. Thus,
 pixels with the surface brightness $X>6.4$ are easily detectable as more than $5\sigma$ deviations on top of the mean level. \alerta{To get 100,000 photons
 in the 0.5--2.0 keV band the Chandra \mbox{ACIS-I} instrument with an effective area\footnote{This is the weighted average effective area over the energy
 range \mbox{0.5--2.0 keV} for a hot gas with $T\sim {\rm \it few}$ keV. The photon rate used to estimate the exposure was also
 computed for the same energy range.} of $\sim$370 cm$^2$ would require an exposure time of $\sim$10 ks for the simulated cluster CL101 with a mass of
 $\sim 1.2 \times 10^{15}~{\rm M_{\odot}}$ and $\sim$150 ks for CL24 with a mass of $\sim 5.0 \times 10^{13}~{\rm M_{\odot}}$, assuming a redshift of $z=0.1$
 for both objects.} Further, we see that the clumps in the image will still be resolved if the size of the pixels is increased by a factor of $\sim 2.5$.
 This reduces the required number of photons, we now find that pixels with surface brightness $X>10.25$ can be used for masking,
 using $f_{\rm cut} = 2.5~{\rm and}~3.0$, even if there are 10,000 counts for the whole image.

\subsection{Results: Bias between X-ray and SZ pressure profiles}
\label{sec:results_bias}
\begin{figure}
 \hspace{-0.5mm}
  \centering
  \includegraphics[width=0.475\textwidth]{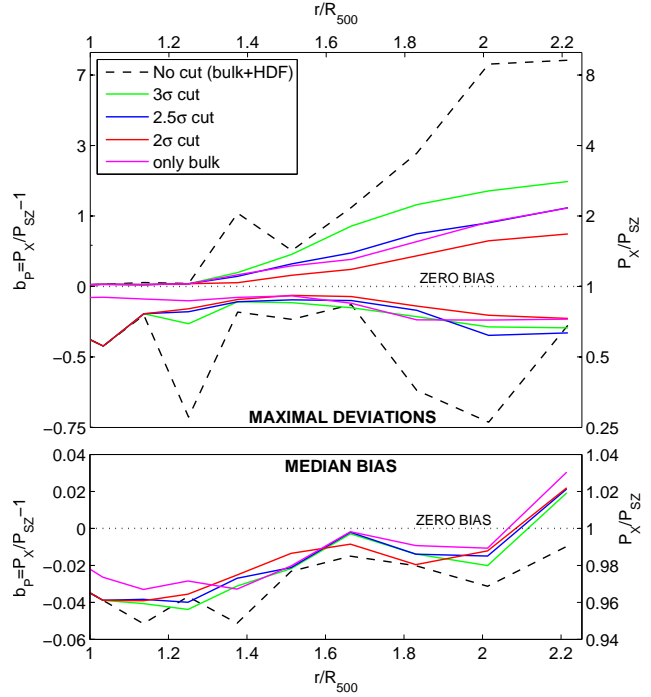}
  \caption{Reduction of the bias, $b_P \equiv P_X/P_{SZ}-1$, in the pressure profiles by the introduction of various cuts in surface brightness
 (applied to the X-ray images), at $r \geq R_{500}$. {\it Upper panel:} \alerta{Maximal deviations (or sample outliers)} for both relaxed and unrelaxed clusters
 in the NR simulation. {\it Lower panel:} Median bias profiles for each of the above cuts. \alerta{Note that in both panels, the left axis shows \mbox{$b_P=P_X/P_{SZ}-1$}
 while the right axis displays \mbox{$P_X/P_{SZ}$}; the dotted line, `ZERO BIAS' indicates \mbox{$b_P=0$} or $P_X/P_{SZ}=1$.}} 
  \label{fig:various_cuts}
\end{figure}

Fig. ~\ref{fig:rad_profile}(a) shows the median bias profile of the relaxed clusters in the NR run.
 The profiles with and without HDF components (the 3-D cut) are shown by the thick solid blue and thick dashed magenta lines,
 respectively. Thick dot-dashed black line shows the median profile with the exclusion of the bright clumps beyond
 $R_{500}$ in the projected map (the 2-D cut with $f_{\rm cut}=2.5$). The sample outliers are indicated by the thin lines of similar type.
 Note that this bias is computed from the composite spectra corresponding to the actual distribution of temperature and density values
 within each shell (see section \ref{sec:xspecmethod}), and not from the idealised lognormal distribution.
 We find the median bias computed from the relaxed clusters
 to be within $0\%$ to $+2\%$ at $r < 0.5 ~R_{500}$. This then decreases to negative values $\sim -6\%$ at $r=R_{500}$. The bias
 profiles remain unaffected by the HDF component (due to small fluctuations) within $R_{500}$, producing almost identical profiles
 with and without HDF. Beyond $R_{500}$, the median bias profiles from `only bulk' follow the same trend as `bulk+HDF',
 but are smoother and are limited to within $-2.5\%$ to $+2\%$. With the 2-D cuts, the median bias profile is also smoother
 and agrees with the bias profile after the 3-D cut (see the lower panel of Fig. \ref{fig:various_cuts}).

We find that the unrelaxed clusters display a larger negative bias and larger scatter, both from shell to shell and across the
 sample at a given value of $r/R_{500}$.
 The cluster-to-cluster scatter increases with radius. At $r=R_{500}$, the scatter
 is less than $3\%$ and $13\%$ for relaxed and unrelaxed clusters, respectively, for the NR run. At larger radii ($r \sim 2 ~R_{500}$), the scatter is highly
 skewed in the positive direction, with some shells showing $b_P \gg 0$ (see Fig. \ref{fig:various_cuts}), from the presence of
 dense clumps. After the 3-D cut, the scatter in the pressure bias drops to less than $7\%$ within $R_{500}$ even in unrelaxed clusters.

 Fig. ~\ref{fig:rad_profile}(b) shows the `measured' values of the X-ray and SZ pressures w.r.t. the {\it median} pressure in concentric
 shells at various radii. \alert{Since the value of $P_{SZ}$ actually corresponds to the average electron pressure}, the fact that there is a bias
 even in the SZ profiles implies that the mean and the median values of pressure fluctuations are different. This is true for any
 asymmetric distribution such as the near-lognormal distribution of pressure fluctuations seen in
 Fig. \ref{fig:distrP}. The plot shows that the `measured' SZ and X-ray pressures are largely similar, the differences between them are smaller than
 differences in the mean and median values of pressure fluctuations.

In Fig. ~\ref{fig:rad_profile}(c) and (d), we also plot the bias in the density and spectroscopic temperature values as would be `measured' from X-ray observations.
As seen from equation \ref{eqn:estimate} the best-fitting values of $n_e$ are biased in the positive direction, while the best-fitting $T_X$ is biased
 negatively. For the relaxed clusters in the NR simulation, the median bias in gas density increases
 slowly to $+3.5\%$ at $R_{500}$ but can reach $\sim +40\%$ at larger radii. Interestingly, the exclusion of bright clumps reduces the median bias to
 +6.5\% at $r=2 ~R_{500}$ and makes the bias profile considerably smoother. Beyond $R_{500}$ the median bias profiles in gas density obtained after
 both 3-D and 2-D cuts are quite similar. The median bias in temperature takes similar values up to $r\sim 0.8 ~R_{500}$ as the density bias,
 but in the negative direction, reaching $\sim -14\%$ at $r=2 ~R_{500}$.
 We find that the median bias in temperature reduces only slightly even after removing the HDF, because the removal of the HDF components only affects the
 density distribution, but does not directly limit the temperature fluctuations. This is because, there is almost zero correlation between density and temperature
 fluctuations (see Fig. ~\ref{fig:rad_profile_others} (b)) in the ICM at $r>R_{500}$. At the same time, the sample outliers producing a negative bias in
 temperature reduce to nearly a third of the original value after the 2-D and 3-D cuts (as the \alert{very} dense clumps are cold).

In Fig.~\ref{fig:various_cuts} we compare the bias profiles in pressure, in the cluster outskirts ($r > R_{500}$),
 for the various cuts in surface brightness applied in concentric annuli on the projected images
 of the NR simulations. We also plot for comparison the results from the `HDF+bulk' (No cut) and `only bulk' (3-D cut).
 We find that application of more stringent cuts (specified by lower values of $f_{\rm cut}$) reduces the scatter in the bias profiles. 
Exclusion of the bright clumps in the cluster outskirts ($r > R_{500}$) reduces the outliers (on the positive side) in the bias ($b_P$) profiles,
 by more than an order of magnitude, from $+8.2$ to $+1.8$, $+1.2$, and $+0.7$ respectively, at $2.2~R_{500}$, for $f_{\rm cut}$ = 3.0, 2.5 and 2.0.
 For $f_{\rm cut}=2.0$ the excluded clumps have
 surface brightness between $\sim$ 9 -- 13 times that of the average value (across the inner to the outer annuli). For $f_{\rm cut}=2.5$ this is 13--18; while for
 $f_{\rm cut}=3.0$ these numbers are 20--28. It is clear that such exclusions would remove only the brightest regions associated with high density
 clumps, as shown in Fig. \ref{fig:mask} for the unrelaxed cluster CL101 in the NR run. On the negative side (i.e. outliers with $b_P < 0$), the various cuts do not
 affect the outliers as much as \alert{for the positive side}. We see from the bias profiles plotted in the upper panel of Fig. \ref{fig:various_cuts}
 that the 2-D cut specified by the
 value $f_{\rm cut} = 2.5$ produces similar results to the 3-D cut. A more stringent cut ($f_{\rm cut} = 2.0$) may further eliminate the intrinsic brightness
 fluctuations associated with inhomogeneities in the diffuse ICM, while a more relaxed cut ($f_{\rm cut} = 3.0$) does not remove all the visibly bright clumps. In
 the lower panel of Fig. \ref{fig:various_cuts} we indicate the median values of the bias for the 16 clusters; these are again quite robust to the application of
 the various 2-D and 3-D cuts.

\begin{figure}
  \centering
  \includegraphics[width=0.48\textwidth]{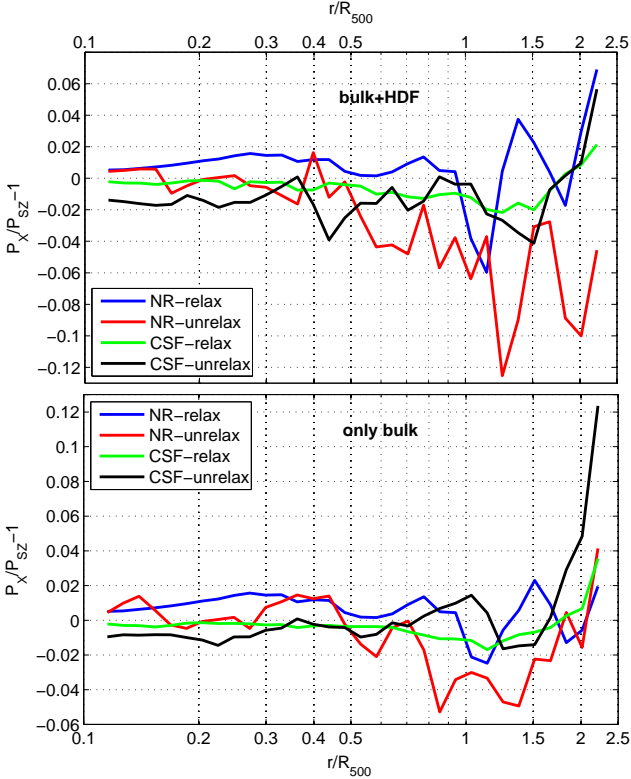}
  \caption{The bias, $b_P(r) \equiv P_X/P_{SZ}-1$, in pressure profiles for a sample of 16 clusters (both relaxed and unrelaxed) in the NR and CSF simulations.
  For the figure in lower panel, labelled as `only bulk', the bias is computed after excluding contributions from the high density fluctuations (HDF)
 associated with dense clumps.}
  \label{fig:all_runs}
\end{figure}

 Finally, we compare the bias in the median pressure profiles, $b_P \equiv P_X/P_{SZ}-1$, for both relaxed and unrelaxed clusters in simulations with \alert{very different}
 input physics. \alert{As expected, we find more clumping in the CSF simulations when compared to the NR simulations; the bias profiles in the CSF runs also display
 significantly larger scatter from shell to shell (over the same radial range) due to the presence of some very dense clumps.}
 Fig. \ref{fig:all_runs} shows the median bias profiles to be within $-4\%$ to $+2\%$ for $r<0.5 ~R_{500}$ and within $-6\%$ to $+2\%$ for $0.5 ~R_{500} < r < R_{500}$
 for relaxed and unrelaxed clusters in both NR and CSF simulations.
At $r\approx 1.2 ~R_{500}$, the bias reaches $-13\%$, arising from the presence of infalling
 clumps, but reduces to less than $-5\%$ when the HDF component is removed. In both the NR and CSF simulations the bias increases beyond $r\gtrsim 2~ R_{500}$.

\alert{\alert{More importantly, we find that the pressure bias is limited to within $\pm$15\% for any of the relaxed simulated
 clusters in both CSF and NR runs of the simulation inside $R_{500}$.} Also, the results on median bias profiles are quite similar (see Fig. \ref{fig:all_runs}) across
 very different input physics used in the NR and CSF runs. This suggests that our results are robust to possible uncertainties in the implementation
 of physical processes in the hydro codes.}

\section{\alert{Comparison with current and future measurements}}
\label{sec:compare}
High-resolution observations of galaxy clusters jointly in X-rays and millimetre/sub-millimetre bands provide independent measures of the thermal pressure
 profiles through a deprojection analysis. In this work we provide an estimate of the expected bias in the pressure profiles derived from the X-ray and microwave
 bands taking into account both temperature and density fluctuations in the ICM, as well as the correlation between them. We also predict biases in three-dimensional
 density and temperature structure derived from X-ray observations. The bias in density depends on the ability (determined by the photon statistics and X-ray background)
 to mask out X-ray emitting gas clumps in the outskirts of galaxy clusters. \alert{Our estimates from section \ref{sec:clumpy} seem to suggest that X-ray observations would
 be able to identify these clumps (if resolved) in clusters up to moderate redshifts \mbox{($z \sim 0.1$)}.}
 
Accurate measurements of the bias as a function of cluster radius will allow us to probe the inhomogeneity and the nature of perturbations (pressure vs. entropy)
 in the ICM. Our work is also relevant for the use of galaxy clusters in cosmological studies. A robust characterisation of biases between X-ray and SZ observations is
 critical for constraining cosmological parameters through measurements of the angular diameter distances, $d_A$, at various redshifts
 \citep{2006ApJ...647...25B, 2010PhRvD..82h1301K}. The prediction of the bias in density profiles is also important for the determination of the baryon budget
 in clusters \cite[e.g.,][]{Simionescu,2012arXiv1211.1695V}.

Our results appear to be consistent with the bias ($-10\%$ to $-40\%$, from the cluster centre to $r = 0.8~ R_{500}$) in the pressure profiles seen
 in the Coma cluster \citep{2012arXiv1208.3611P}, using data from XMM-Newton and Planck observations. Coma is an unrelaxed cluster and
 the bias outliers seen in the small sample of unrelaxed simulated clusters range from $-15\%$ to $+20\%$ and $-60\%$ to $+20\%$, for the NR and CSF simulations
 respectively, in the same radial range. The current results on the ratio $Y_{SZ}/Y_X$ appear to be mixed and inconsistent with each other.
On the one hand, recent observations from the CARMA \citep{2012NJPh...14b5010B} indicate that this ratio is consistent with unity, while other measurements
\citep[for e.g.,][]{2012arXiv1202.2150R, 2011ApJ...738...48A} are quite different from the biases derived from our simulations. In our view this is likely due
 to some underlying systematic uncertainties associated with the interpretation of the current observations
 \citep[see][for a list of possible systematic errors]{2006ApJ...647...25B}.

Recently, \citet{Battaglia} estimated biases in the gas mass fraction, $f_{\rm gas}$, derived from X-ray observations, using a set of cosmological cluster
 simulations. Our results on the bias in the derived gas density from X-ray data due to clumpiness are consistent with their findings. They estimate the
 contribution of clumping to $M_{\rm gas}(<R_{200})$ to be $10--20\%$ while our results indicate a {\it median} bias of $16\%$ in the density at the
 same radius.

\section{Conclusions}
The main conclusions of this paper are summarised below:

\begin{itemize}

 \item We first investigate the dependence of bias between the X-ray and SZ pressures by fitting the X-ray emission from an idealised hot gas having
 a lognormal distribution of density and temperature values. Our results demonstrate that the bias is expected to be small (within $\pm 10\%$), as long as the magnitude of
 density and temperature fluctuations is sufficiently small ($\sigma_{T},~\sigma_{ne} < 0.45$) and their correlation does not take large negative values ($\xi > -0.3$), for
 median temperatures, $T_{\rm median} \geq 3$ keV. However, the bias may be somewhat larger ($\sim 15$\%) for clusters with lower temperatures ($T_{\rm median} \lesssim$ 1 keV).
 We provide a fitting form to predict the bias as a function of the properties of fluctuations of the ICM for $T_{\rm median} \geq 3$ keV and plot
 bias contours in the ($\xi$, $\sigma$) plane that may be used to probe the properties of fluctuations from a given observed ratio of $P_X/P_{SZ}$.
At lower temperatures ($T_{\rm median} < 3$ keV), the bias is also found to be a function of the metal abundance in the gas.

 \item We then use a sample of 16 simulated clusters to study properties of the ICM fluctuations. The clusters in the
 simulations show an increase in the amplitude of both density and temperature fluctuations with radius.
The fluctuations of temperature and density are mostly negatively correlated within $r<R_{500}$. This correlation is close to zero at higher radii.
\alert{We compute $P_X$ and $P_{SZ}$ profiles using the ICM from simulated clusters as would be measured from X-ray and SZ observations.}
Both $P_{SZ}$ and $P_X$ are biased with respect to the median pressures values. For $P_{SZ}$ this bias is purely due to the asymmetric (near lognormal)
 distribution of pressure fluctuations; for $P_X$, in addition to the bias from asymmetry, there is also a bias arising from fitting a single temperature
 model to the multi-component emission spectrum.

 \item We show that the median bias, $\mbox b_P(r) \equiv P_X(r)/P_{SZ}(r) - 1$, between the pressure estimated from X-ray and SZ observations is small and lies
 within -6\% to +2\%, up to $R_{500}$, even for unrelaxed clusters. The scatter in \alert{the} pressure bias is significantly smaller for relaxed clusters ($<0.03$)
 than unrelaxed clusters ($<0.13$) at $r<R_{500}$; however, it becomes large and positively skewed at $r\sim 2 ~R_{500}$.  For non-relaxed
 clusters, there is a noticeably higher scatter in the bias values from shell to shell. The outliers responsible
 for the scatter in bias can be reduced by almost an order of magnitude just by using a simple method of excluding the brightest clumps
 (as permitted by reasonably good photon statistics) from the cluster outskirts ($r>R_{500})$, for both relaxed and unrelaxed clusters.

\item \alert{Finally, our results on the median values of pressure bias between X-ray and SZ observations are not significantly different in the two types of simulations with
 different input physics: (i) Non-radiative (NR) runs and (ii) Cooling + star formation (CSF) runs. Therefore these results, (i.e. a small bias within $R_{500}$)
 may be considered to be robust with respect to possible uncertainties in the physical implementations in the hydro codes.}

\end{itemize}

\section*{Acknowledgements}
\alert{The authors would like to thank the referee for his/her suggestions and comments which have been useful in improving the presentation of this work.}
 IZ, EC, AK, DN and EL thank KITP for hospitality during the workshop ``Galaxy Clusters: the Crossroads of
Astrophysics and Cosmology''(2011). The work was supported in part by the Division of Physical Sciences of the RAS
 (the program ``Active processes in galactic and extragalactic objects'', OFN-17). AK was supported in part by NSF
 grants AST-0807444 and AST-0904484, NASA grant NAG5-13274, and by the Kavli Institute for Cosmological Physics at the University of
 Chicago through the NSF grant PHY-0551142 and PHY-1125897 and an endowment from the Kavli Foundation. EL was
 supported in part by NASA Chandra Theory grant GO213004B. DN acknowledges support from NSF grant AST-1009811, NASA ATP
grant NNX11AE07G, NASA Chandra Theory grant GO213004B, Research Corporation, and Yale University. This work was
 supported in part by the facilities and staff of the Yale University Faculty of Arts and Sciences High Performance
Computing Center. The cosmological simulations used in this study were performed on the IBM RS/6000 SP4 system (copper)
 at the National Center for Supercomputing Applications (NCSA).

\bibliographystyle{mn2e}

\bsp

\label{lastpage}

\end{document}